\documentclass[acmlarge]{acmart}

\usepackage{caption}
\usepackage{graphicx}
\usepackage{xspace}
\usepackage{cleveref}
\usepackage{lipsum}
\usepackage[normalem]{ulem}
\usepackage{enumitem}
\usepackage{geometry}
\usepackage{amsmath}
\setlist[itemize]{noitemsep, topsep=0pt}
\usepackage[utf8]{inputenc}
\usepackage[T1]{fontenc}
\usepackage{adjustbox}
\usepackage{subfigure}
\usepackage{multirow}
\usepackage[normalem]{ulem}

\usepackage{makecell}
\usepackage{mdframed}
\usepackage{soul}
\usepackage{hyperref}
\hypersetup{
    colorlinks=true,
    linkcolor=blue,
    filecolor=magenta,
    urlcolor=cyan,
}

\usepackage{siunitx}
\usepackage{tikz}

\makeatletter
\DeclareRobustCommand\onedot{\futurelet\@let@token\@onedot}
\def\@onedot{\ifx\@let@token.\else.\null\fi\xspace}

\makeatother

\AtBeginDocument{%
  \providecommand\BibTeX{{%
    \normalfont B\kern-0.5em{\scshape i\kern-0.25em b}\kern-0.8em\TeX}}}

\begin{document}
\title[]{Exploring Large Language Model as an Interactive Sports Coach: Lessons from a Single-Subject Half Marathon Preparation}


\author{Kichang Lee}
\email{kichang.lee@yonsei.ac.kr}
\affiliation{%
  \institution{Yonsei University}
  \country{Republic of Korea}
}


\begin{abstract}
Large language models (LLMs) are increasingly deployed as domain-adaptive assistants, yet their prospective role as longitudinal, everyday \emph{virtual coaches} remains underexplored. This single-subject, two-month case study examines an LLM-guided preparation for a half marathon (July 25–September 28, 2025). The participant interacted with a general-purpose LLM primarily via text, uploading training logs and occasional screenshots from consumer apps. We analyze the coaching interaction along three roles—\emph{planner, explainer, motivator}—and quantify outcomes with session- and week-level metrics. Empirically, running performance improved from sustaining only $\sim$2\,km at $7'54''$/km to completing 21.1\,km at $\sim$$6'30''$/km on race day; cadence rose from $\sim$160\,spm toward 170–180\,spm; pace–heart rate relationships shifted favorably; and a derived efficiency index (speed/HR) exhibited upward trends in 7- and 28-day medians. While causal attribution is limited absent a control condition, these convergent indicators document meaningful progression under LLM-guided planning.
Beyond outcomes, we surface recurring limitations that bound present-day LLM coaching: (i) \textbf{modality and latency gaps}—no in-run pose/HR context and predominantly \emph{post hoc} dialogue; (ii) \textbf{outbound feedback constraints}—text-only guidance for embodied skills; (iii) \textbf{motivation asymmetry}—support largely \emph{user-initiated}, weakest when most needed; and (iv) \textbf{model specialization and safety}—limited persistence, guardrails, and personalization. We outline design requirements for next-generation systems: persistent athlete models with safety-aware progression logic; multimodal, on-device sensing for near–real-time cues; audio/haptic/visual feedback loops; proactive, consented motivation scaffolds; and privacy-preserving personalization (e.g., federated learning, test-time adaptation). Taken together, this study offers grounded evidence and a concrete design agenda for evolving LLMs from retrospective advisors toward closed-loop, safe, and personalized coaching companions.
\end{abstract}
\settopmatter{printfolios=true} 
\settopmatter{printacmref=false} 
\renewcommand\footnotetextcopyrightpermission[1]{} 
\maketitle

\section{Introduction}
\label{sec:intro}
Large language models (LLMs) have rapidly evolved from general-purpose chatbots into domain-adaptive assistants that can reason over goals, constraints, and evolving context. Across diverse sectors, their ability to synthesize expert knowledge, generate actionable plans, and support reflective decision-making has been repeatedly demonstrated. Recent deployments in safety-critical domains such as healthcare~\cite{yang2024drhouse} and in cyber-physical operations such as building HVAC optimization~\cite{lee2025llm} underscore an emerging pattern: when paired with user-specific constraints and feedback, LLMs can function not merely as information-retrieval tools but as \emph{interactive, situated collaborators}. This observation motivates an inquiry into their prospective role in everyday human performance settings—particularly in endurance training, where adherence, progressive overload, and recovery management are pivotal yet often under-institutionalized for non-elite participants.

Sports-related applications of LLMs are growing, from rule clarification and exercise-prescription summaries to lightweight coaching embedded in consumer platforms. However, current practice is still dominated by \emph{episodic} interactions: one-off queries about form cues (e.g., “how to run a tempo session?”), nutrition heuristics (e.g., “what to eat before a long run?”), or injury-prevention checklists. While such micro-interventions are useful, they fall short of what human coaches provide—a \emph{longitudinal} understanding of the athlete’s objectives, history, constraints, fatigue, and behavioral patterns, coupled with timely adjustments and motivational scaffolding. In short, despite surging enthusiasm, we lack systematic analyses of whether, when, and how an LLM can sustain the role of a virtual coach over \emph{weeks to months}, continuously ingesting user states and outcomes to recommend, adapt, and motivate.

Running provides a compelling testbed for this question. As one of the most accessible forms of exercise, running requires minimal equipment, is logistically simple, and enjoys widespread participation. Yet for recreational runners, performance plateaus, overuse symptoms, and program non-adherence are common. Typical pain points include uncertainty about periodization (how to phase base, tempo, interval, taper), pacing (how to regulate intensity within and across sessions), and recovery (how to modulate load in the presence of soreness or life constraints). Traditional remedies—books, generic templates, or sporadic forum advice—offer limited personalization and little real-time reactivity to an individual’s fluctuating capacity and schedule. A dialogic agent that can track goals, read training logs, and reason about trade-offs could therefore address an unmet need: \emph{dynamic, context-aware guidance} for non-elite runners.

This paper presents a two-month, single-subject case study probing the feasibility and limitations of LLM-assisted half-marathon preparation. From July~25, 2025 through race day on September~28, 2025 (Incheon Songdo International Marathon, half course), the author engaged an LLM (hereafter, “the agent”) as an interactive coach. The author supplied the agent with evolving constraints (e.g., weekly availability, session durations), short-term goals (e.g., target paces for upcoming workouts), and a longitudinal objective (half-marathon completion at a specific pace). In return, the agent proposed periodized plans, made session-level adjustments in light of fatigue and scheduling changes, provided educational rationales to support adherence, and offered motivational prompts when setbacks occurred. The study’s intent is \emph{not} to claim superiority of LLM coaching over human coaching, but to document, in situ, the interaction patterns, decision points, and athlete outcomes that can inform future, larger-scale investigations.

Empirically, the author began the training block with modest capacity, able to sustain only a short continuous effort (approximately $2\,\mathrm{km}$) at a pace of $7'54''/\mathrm{km}$, equivalent to about $7.59\,\mathrm{km/h}$. Over the subsequent weeks—following agent-guided progressions in easy mileage, strides, aerobic intervals, and long-run structure, together with recovery and strength suggestions—the author completed the target event, covering $21.1\,\mathrm{km}$ at approximately $6'30''/\mathrm{km}$ ($\approx 9.23\,\mathrm{km/h}$). While a single case cannot establish causality, the magnitude and \emph{specificity} of improvement, occurring under explicit schedule and fatigue constraints negotiated through dialogue, provide a concrete probe of where LLM-based coaching presently helps or hinders.

Our objective is deliberately pragmatic: we document how an LLM was used in practice during a two-month half-marathon build and, from that lived experience, identify concrete limitations and sketch what future sports-training applications will need. Rather than offering a comprehensive HCI analysis of “interaction mechanics,” we ground claims in observed use: where the agent’s planning, explanations, and motivational support helped; where they fell short; and how those shortcomings manifested amid real scheduling, fatigue, and recovery constraints.

Based on this experience, we emphasize three key limitations of current LLM-based training assistance. \textbf{First}, while the agent effectively produced a long-term plan and kept the user aligned with high-level goals, it lacked access to the runner’s physical state \emph{during} training sessions (e.g., heart rate, cadence, posture). Without additional modalities from wearable or motion sensors, the LLM cannot provide accurate in-situ adjustments or semi–real-time feedback; future systems should integrate sensor streams and calendar information to deliver context-aware, adaptive coaching. \textbf{Second}, the communication channel remained text-only, constraining the expressiveness of feedback and instruction. Subtle elements such as form corrections or pacing cues are difficult to convey in text alone; multimodal feedback (visual, auditory, haptic) will be required to make coaching actionable in the moment. \textbf{Third}, even the best-designed plans are meaningless if athletes do not execute them. The case study underscored that sustained motivation and adherence are the hardest parts of training. While the LLM provided encouragement when prompted, future applications should embed more robust mechanisms for long-term engagement (e.g., gamification, progress visualization, adaptive motivational interventions).

In light of these limitations, we outline functional requirements for next-generation LLM coaching systems: (i) integration with multimodal sensing devices to capture real-time physical states; (ii) mechanisms for semi–real-time feedback delivery during exercise, going beyond text to richer modalities; and (iii) motivational frameworks that sustain long-term adherence, ensuring that plans translate into consistent training behavior. These requirements, distilled from practical use in this single-subject study, provide a roadmap for evolving LLMs from static advisors into truly interactive, effective virtual coaches.

The present work makes three contributions. \textbf{First}, it offers a richly annotated, longitudinal account of LLM-assisted endurance preparation, articulating concrete points where dialogic planning and explanation supported adherence and learning—and where they did not. \textbf{Second}, it proposes an analysis framework for \emph{virtual coaching roles} (planner / explainer / motivator) and links these roles to measurable process and outcome variables relevant to recreational sport. \textbf{Third}, it distills design implications for next-generation LLM coaching systems that integrate real-time sensing, safety-aware progression logic, and reflective prompts to sustain motivation. Collectively, these contributions provide groundwork for controlled studies with larger cohorts, stronger baselines (e.g., certified coach programs, adaptive training apps), and objective endpoints (e.g., field tests and validated readiness scales).

In summary, this single-subject case study does not claim generality by itself; rather, it \emph{constructively problematizes} what effective, safe, and motivating LLM coaching should entail for everyday athletes. By tracing how an agent and runner co-adapted plans under real constraints to reach a specific event goal within two months, we illuminate both the promise and the boundaries of current LLMs as virtual coaches, motivating a research agenda toward real-time, personalized, and ethically responsible AI support in recreational endurance training.

\section{Related Work}

\subsection{Large Language Models in Applied Domains}
Large language models (LLMs) have progressed from general conversational systems to domain-adaptive assistants that scaffold concrete workflows. In healthcare, a recent JAMA systematic review synthesizing 126 studies finds that evaluations overwhelmingly emphasize examination- or vignette-style testing rather than end-to-end clinical use with real patient data; moreover, fairness, safety, and deployment readiness are inconsistently addressed~\cite{bedi2025testing}. Subsequent surveys echo these gaps, noting that while LLMs show competence on diagnostic questions and summarization tasks, prospective evaluations embedded in clinical workflows remain rare~\cite{bedi2025testing}.

Beyond medicine, applied work targets cyber-physical settings where language interfaces can bridge expert tooling. In building energy modeling, \emph{EPlus-LLM} fine-tunes a text-to-specification pipeline that translates natural-language requirements into EnergyPlus models and automates simulations, improving efficiency for routine modeling tasks~\cite{jiang2024eplus}. Complementary efforts use LLMs to normalize and map Energy Efficiency Measures across heterogeneous taxonomies, facilitating analytics and retrofit planning~\cite{khanuja2024can}. Across these domains, LLMs reduce friction at \emph{task boundaries} (e.g., specification, documentation, schema translation) and provide integration scaffolding for existing tools.

However, the evidence base still skews toward short, controlled trials, ablation-style demonstrations, or case reports without sustained outcome tracking~\cite{bedi2025testing,jiang2024eplus,khanuja2024can}. In contexts where progress accrues over weeks to months—such as endurance training—the absence of \emph{longitudinal} evaluations constrains claims about real-world effectiveness, safety, and adherence. This gap motivates our event-focused, two-month case study that treats an LLM as a training assistant and records both process (plans, adjustments) and outcomes (pace changes, race completion).

\subsection{Conversational Agents for Health and Fitness}
Conversational agents for physical activity and lifestyle change predate the LLM era and have been synthesized by mixed-methods and quantitative reviews. Luo et al.\ report feasibility and user acceptability for agents promoting physical activity, while highlighting repetitive content, variable study quality, and sparse long-term follow-up~\cite{luo2021promoting}. A meta-analysis by Singh et al.\ covering chatbot interventions across behaviors (physical activity, diet, sleep) finds small-to-moderate efficacy, with notable heterogeneity and limited durability of effects~\cite{singh2023systematic}. Aggarwal et al.\ similarly observe short study durations prioritizing usability metrics over rigorous, longitudinal outcomes~\cite{aggarwal2023artificial}.

LLM-era prototypes push toward richer counseling and personalization. \emph{GPTCoach} examines LLM-delivered coaching aligned with health-coaching practices and tool use over wearables; however, its evaluation is short-term and centers on experience measures rather than month-scale performance endpoints~\cite{jorke2025gptcoach}. In parallel, \emph{Coaching Copilot} blends an LLM chatbot with a human coach to support reflection and behavior change, demonstrating promise for engagement over a two-week window in a non-athletic leadership context~\cite{arakawa2024coaching}.

Taken together, prior work establishes that conversational agents—including LLM-powered systems—are usable, can convey evidence-based guidance, and may improve short-term engagement. Yet across health and fitness, \emph{longitudinal, athlete-level} evaluations that trace coaching interactions, safety constraints (e.g., load progression), and objective performance trajectories (e.g., pacing across micro-/mesocycles, event completion) remain scarce~\cite{luo2021promoting,singh2023systematic,aggarwal2023artificial,jorke2025gptcoach}. Our study addresses this gap by documenting two months of LLM-assisted half-marathon preparation with concrete process and outcome measures, complementing short-horizon usability studies with an event-focused, real-world training narrative.

\subsection{LLMs for Marathon Training and Commercial AI Tooling in Running}
Beyond academic prototypes, the application of LLMs to endurance running is emerging in the wild through grassroots use and creator-made agents. Public narratives document runners using ChatGPT to generate half- and full-marathon plans tailored to their schedules and constraints, then refining these plans iteratively as training progresses~\cite{medium_build_plan_chatgpt}. Community discussions similarly report leveraging LLMs for workout structure, nutrition prompts, and race-week checklists, indicating early adoption as a \emph{DIY virtual coach} among recreational athletes~\cite{facebook_chatgpt_marathon_post}. In parallel, individuals have published custom GPT-based coaches on consumer platforms (e.g., “Marathon GPT,” “Marathon Prep Coach”), packaging prompt engineering and domain heuristics into reusable coaching experiences for non-experts~\cite{gpt_marathon_gpt,gpt_marathon_prep,gpt_marathon_training_plan}.

Commercial platforms are also introducing AI features that intersect with training guidance and performance understanding. On Strava, \emph{Athlete Intelligence} analyzes completed workouts to surface personalized insights, while \emph{Performance Predictions} provides distance-specific estimated finish times to guide pacing and preparation; recent product cycles further highlight AI- and ML-driven route planning and leaderboard-integrity systems~\cite{strava_athlete_intel,strava_perf_predictions,verge_strava_ai_routes,press_strava_aug2025}. Wearable ecosystems expose related coaching primitives. Samsung Health’s running coach on Galaxy Watch evaluates an athlete’s level using heart rate, pace, and estimated $\mathrm{VO_2max}$, then recommends a personalized plan and progression; recent Galaxy AI updates expand on-device analytics and structured workout routines for endurance preparation~\cite{samsung_running_coach_latin,samsung_running_coach_lb,samsung_galaxy_ai_watch}.

Despite rapid adoption, \emph{evaluation remains largely short-horizon and anecdotal}. Creator posts and community agents provide persuasive use stories but rarely report athlete-level outcomes over mesocycle timescales, and commercial announcements emphasize capability launches rather than prospective, longitudinal performance or safety endpoints~\cite{medium_build_plan_chatgpt,facebook_chatgpt_marathon_post,strava_athlete_intel,strava_perf_predictions}. This evidentiary gap motivates our two-month, event-focused case study, which documents both process (dialogue-driven plans and adjustments) and outcomes (pace trajectories and race completion) under real-world constraints.

\section{Study Design}
\label{sec:design}
We describe the study design for a two-month, event-focused investigation of LLM-assisted half-marathon preparation.

\subsection{Objective and Research Questions}
The objective was to examine whether an LLM-assisted training protocol could enable a novice runner to complete a half-marathon (21.1\,km) within a nine-week build \emph{and} to characterize the decision-level interactions that occurred in practice. We formalize three research questions:

\begin{itemize}
  \item[\textbf{RQ1:}] Over a two-month build, how did key training/process metrics change under LLM-assisted planning?
  \item[\textbf{RQ2:}] In what concrete situations did the LLM help or hinder decisions (e.g., pace setting, session selection, load progression)?
  \item[\textbf{RQ3:}] What functional gaps emerged regarding safety, progression, and in-situ feedback?
\end{itemize}

\subsection{Participant and Context}
The participant was a single male, age~25, height~174\,cm ($\approx$\,5'7"), baseline body mass~85\,kg, with minimal endurance background (occasional weight training; no structured aerobic conditioning). The training period extended from July~25 to September~28, 2025, culminating in the Incheon Songdo International Half Marathon (21.1\,km). During preparation, daily bioimpedance measurements (InBody H20) indicated a $\sim$4--5\,kg body-mass reduction with an estimated $\sim$1\,kg increase in lean mass.

This study adopts an \emph{auto-ethnographic} design: the participant is also the author. The approach enabled fine-grained documentation of lived interaction with an LLM coach but limits generalizability and introduces potential self-report bias; we discuss threats to validity in Section~\ref{sec:threats}.

\subsection{Data Sources}
We relied on three primary data categories:

\begin{enumerate}[label=(\roman*)]
  \item \textbf{Structured training logs.} Each session included date, session type, distance, average pace, and session RPE, with free-text notes on fatigue, soreness, and sleep. Average heart rate (HR) was included when the watch was available; sessions without HR are flagged as missing and excluded from HR-derived metrics (e.g., efficiency index).

  \item \textbf{LLM interaction logs.} The participant used the publicly available \emph{ChatGPT} service for accessibility and familiarity. Screenshots from Nike Run, iPhone Fitness, and Zepp were regularly uploaded to supply structured context (date, distance, pace; HR when recorded). Dialogues covered meso/micro planning, weekly adjustments, explanatory rationales, and motivational support.

  \item \textbf{Milestones and outcomes.} We recorded adherence lapses (missed/truncated sessions), notable fatigue or soreness episodes, and race-day outcomes.
\end{enumerate}

\begin{figure}[h!]
    \centering
    \includegraphics[width=\linewidth]{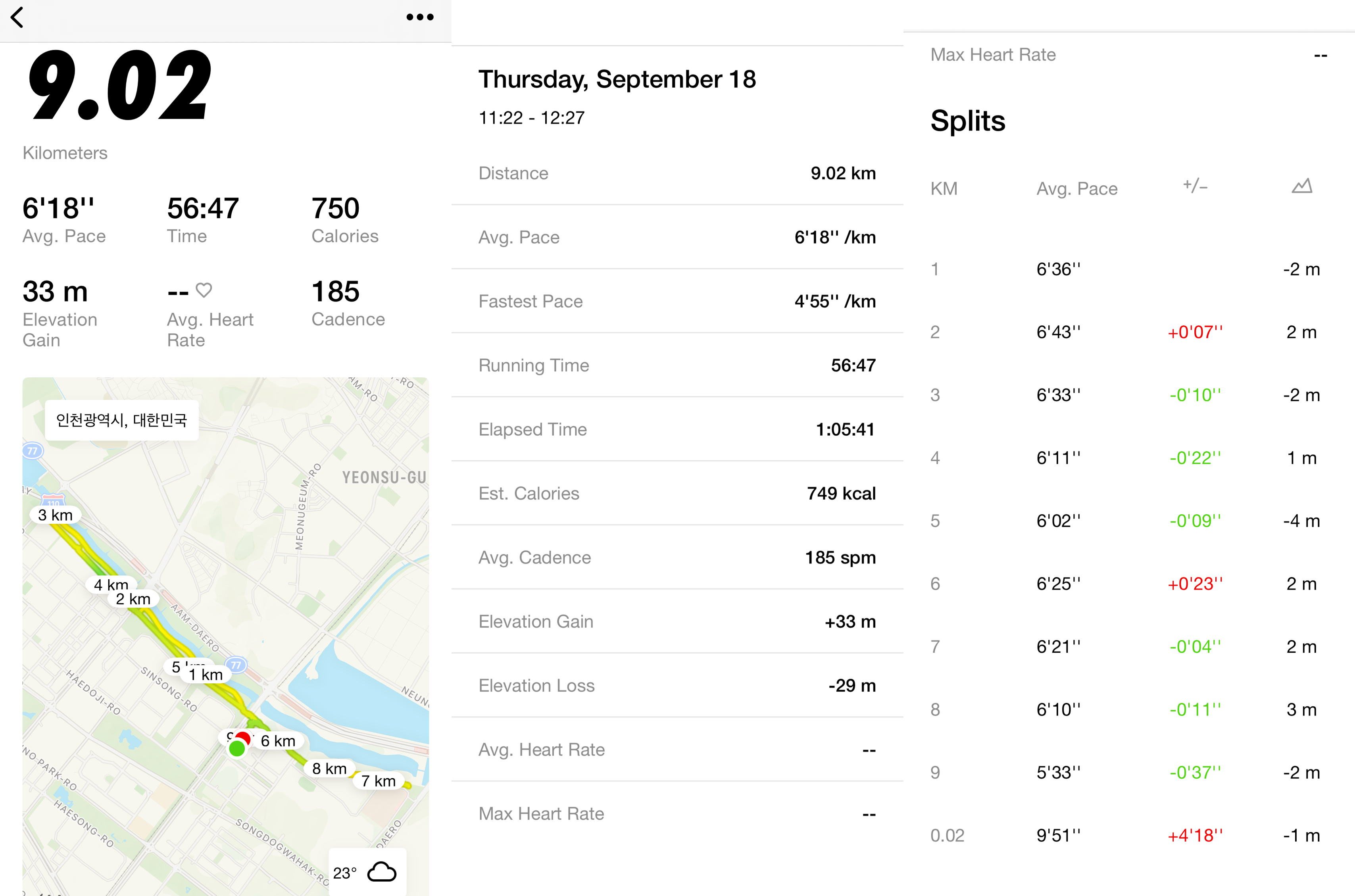}
    \caption{Representative Nike Run Club screenshots. Core fields (date, distance, average pace, route) were consistently available and shared with the LLM.}
    \label{fig:nike}
\end{figure}

\begin{figure}[h!]
    \centering
    \includegraphics[width=\linewidth]{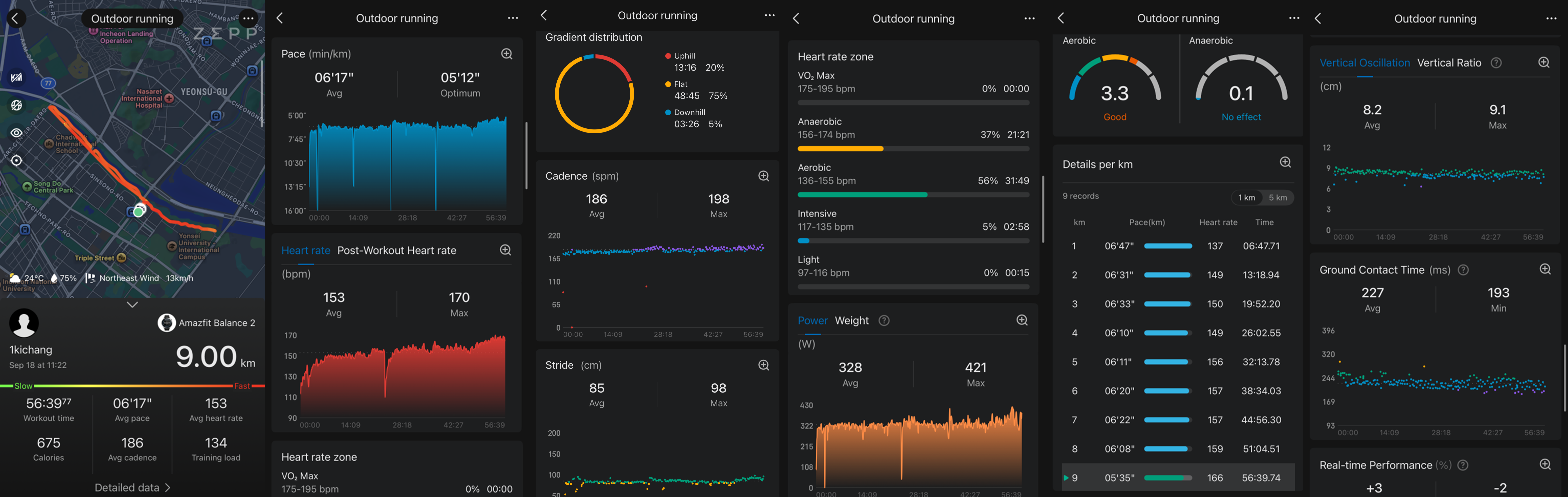}
    \caption{Representative Zepp screenshots paired with Amazfit Balance~2. Compared with Nike Run Club, Zepp offered richer physiological and training signals (continuous HR, cadence, stride, estimated $\mathrm{VO_2max}$, recovery indices).}
    \label{fig:zepp}
\end{figure}

\noindent\textbf{Devices.} Early sessions were recorded on an Apple Watch SE (HR and cadence). Midway, device failure prompted a switch to an Amazfit Balance~2 paired with the Zepp ecosystem; subsequent HR/pace data were collected from this device. The smartphone was an iPhone~16 Pro for app logging and synchronization. Representative screenshots are shown in Figures~\ref{fig:nike} and~\ref{fig:zepp}. Nike Run Club consistently surfaced date, distance, pace, and route, forming the baseline input to the LLM after each session (Figure~\ref{fig:nike}). In contrast, Zepp provided richer training signals via the Amazfit Balance~2—continuous HR, cadence, stride length, estimated $\mathrm{VO_2max}$, and recovery indices—enabling more precise contextualization when these data were uploaded (Figure~\ref{fig:zepp}). The mid-study device change is noted as a potential source of measurement heterogeneity and considered in Section~\ref{sec:threats}.

\begin{table}[h]
\centering
\caption{Training-log schema (core fields).}
\label{tab:training-schema}
\begin{tabular}{ll}
\toprule
Field & Description \\
\midrule
\texttt{date} & ISO date of session \\
\texttt{type} & easy / tempo / intervals / long / rest / strength \\
\texttt{dist\_km} & distance (km) \\
\texttt{pace\_minpkm} & average pace (min/km) \\
\texttt{hr\_avg} & average HR (bpm), if available \\
\texttt{rpe} & 1--10 session RPE \\
\texttt{notes} & free-text remarks (fatigue, soreness, sleep) \\
\bottomrule
\end{tabular}
\end{table}

\subsection{LLM Interaction Protocol}
Interactions with the LLM occurred approximately 4--5 times per week. Inputs included goals, weekly availability, recent logs, and subjective state; outputs included weekly plans, session-level adjustments, and explanatory rationales. We coded messages into three functional roles—planner, explainer, motivator—using the codebook in Table~\ref{tab:interaction-roles}.

\begin{table}[h]
\centering
\caption{Codebook for LLM interaction roles.}
\label{tab:interaction-roles}
\begin{tabular}{lll}
\toprule
Code & Definition & Example \\
\midrule
Planner & micro/meso plan, schedule repair & ``Move tempo to Thu due to fatigue.''\\
Explainer & theory/rationale & ``\emph{Why} taper reduces residual fatigue.''\\
Motivator & adherence/morale & ``Missed runs $\to$ redistribute load.''\\
\bottomrule
\end{tabular}
\end{table}

\begin{figure}[h!]
\centering
\begin{mdframed}[backgroundcolor=gray!5,roundcorner=5pt]
\small
\textbf{Instruction prompt provided to the LLM at study start.}\\[0.5em]
You are my personal running coach. Starting today (July 25, 2025), I will train for a half-marathon (21.1\,km) scheduled on September 28. My goal is to complete the race within 2 hours 15 minutes.

\medskip
Your role is to help me daily with training plans, recovery, nutrition, stretching, log management, and motivation. When I provide training logs or status updates, you should give tailored advice. If I skip training, suggest motivational feedback and compensatory plans.

\medskip
\textbf{Baseline information:}
\begin{itemize}\itemsep2pt
  \item Date of birth: August 11, 2000; Sex: Male
  \item Weight/Height: 85\,kg / 174\,cm
  \item Prior exercise: occasional weight training
  \item Goal: half-marathon completion within 9 weeks, target $<$2h15m
  \item Availability: 4--5 days/week, 30--90 minutes/day
\end{itemize}

Every morning, provide a summary of the training plan. After each workout, I will log my data, and you should return analysis and feedback. Include advice on diet, sleep, and psychological aspects when necessary. Weather and equipment tips are also welcome. Finally, manage the long-term plan by providing weekly summaries.
\end{mdframed}
\caption{Initial instruction (system prompt) supplied to the LLM to standardize its coaching role.}
\label{fig:instruction}
\end{figure}

Figure~\ref{fig:instruction} shows the initial instruction that anchored subsequent interactions. It emulates a human-coach intake, specifying demographic/background details, constraints, and expectations for daily/weekly planning and post-session feedback. Each session thereafter began with an upload of a captured workout summary (Figures~\ref{fig:nike}--\ref{fig:zepp}), after which the model responded within this instruction boundary. This structure provided consistency while making explicit a key limitation of prompt-based coaching: reliance on user-supplied context and the absence of in-run sensing.

\begin{figure}[h!]
\centering
\begin{mdframed}[backgroundcolor=gray!5,roundcorner=5pt]
\small
\textbf{Participant input (July 25, 2025).}\\
I ran 2\,km on the school track to assess my current condition.\\
\quad Average pace: 7'48''/km \quad Cadence: 166\,spm \quad Avg HR: 160\,bpm\\
Please analyze my current state and prepare an overall training plan until the half-marathon on September 28.

\medskip
\textbf{LLM response (excerpt).}\\
\emph{Assessment:} Base pace indicates continuous running capacity; HR $\approx$\,82\% of estimated max suggests near-threshold stress; cadence 166\,spm is slightly below the 170--180\,spm target.\\
\emph{Roadmap (9 weeks):} Weeks 1--3 base building; 4--6 endurance extension (add intervals/tempo); 7--8 pace adaptation; 9 taper. Example week includes easy runs, cadence drills, cross-training/strength, and a progressive long run.\\
\emph{Recommendations:} Nutrition (pre/post), recovery (7--8\,h sleep; 1 full rest day), stretching (dynamic pre, static post), and equipment tips.
\end{mdframed}
\caption{Illustrative participant--LLM exchange showing initial assessment and a condensed nine-week roadmap.}
\label{fig:interaction-sample}
\end{figure}

\noindent\textbf{Role expression in practice.}
Beyond the concise definitions in Table~\ref{tab:interaction-roles}, the agent exhibited:
\emph{(i) Planner}—constructing micro/meso cycles and repairing schedules (e.g., swapping a tempo run after reported fatigue while preserving weekly volume);
\emph{(ii) Explainer}—attaching training-theory rationales (aerobic base, threshold, taper) that increased transparency and adherence; and
\emph{(iii) Motivator}—reframing lapses as load redistribution opportunities and offering simple prompts that supported continuity.

\section{Metrics and Analysis}
\label{sec:metrics}

We quantified outcomes at three levels: \emph{(i) primary performance metrics}, \emph{(ii) process and safety/progression indicators}, and \emph{(iii) decision-level qualitative mappings} to the LLM’s functional roles. Unless noted, all time-series summaries are reported as distance-weighted daily values aggregated to weekly medians with 7-day and 28-day rolling windows.

\subsection{Primary Outcomes}
Primary performance outcomes included improvements in pace (min/km), distance capacity, and completion of the target half marathon (21.1\,km). We report baseline vs.\ post-block comparisons using session- or week-level 

\[
\text{pace}_{\mathrm{min/km}} \;=\; \frac{60}{v_{\mathrm{km/h}}}, 
\qquad
\Delta\text{pace} \;=\; \text{pace}_{\text{baseline}} - \text{pace}_{\text{post}},
\qquad
\%\Delta\text{pace} \;=\; \frac{\Delta\text{pace}}{\text{pace}_{\text{baseline}}}\times 100.
\]

We also summarize \emph{distance capacity} as the longest continuous run completed each week and report race-day pace and finish time as terminal outcomes.

\subsection{Process Metrics}
We tracked exposure and distributional characteristics of training:
\begin{itemize}\itemsep2pt
  \item \textbf{Weekly mileage} (km) and \textbf{long-run distance} (km).
  \item \textbf{Intensity distribution} by session type (easy/tempo/interval/long/recovery/strength) and by HR zones when available (Z1--Z5 based on estimated maximal HR).
\end{itemize}

\subsection{Cardiovascular Efficiency Metric}
To probe aerobic adaptation, we focused on a single, interpretable metric capturing speed–HR coupling.

\paragraph{Efficiency Index (EI).}
\[
v \;=\; \frac{60}{\text{pace}_{\mathrm{min/km}}} \quad (\mathrm{km/h}), 
\qquad
\mathrm{EI} \;=\; \frac{v\;(\mathrm{km/h})}{\mathrm{HR}\;(\mathrm{bpm})}
\quad \big(\mathrm{km\,h^{-1}\,bpm^{-1}}\big).
\]
Higher EI indicates greater speed per unit cardiac cost. We report daily EI together with 7- and 28-day rolling medians to visualize short- and medium-horizon trends in cardiovascular efficiency.

\subsection{Qualitative Mapping to LLM Roles}
Decision-level episodes were coded into the LLM’s \emph{planner}, \emph{explainer}, and \emph{motivator} roles and linked to proximate outcomes (adherence preserved, load capped, soreness reduced). Illustrative examples:

\begin{table}[h]
\centering
\caption{Decision episodes mapped to LLM roles and proximate outcomes.}
\label{tab:decision-episodes}
\begin{tabular}{p{2.2cm}p{2.6cm}p{7.2cm}}
\toprule
Date & Role(s) & Decision \& Outcome \\
\midrule
Aug~12 & Planner & Swap tempo $\rightarrow$ easy after DOMS; weekly mileage preserved; back-to-back intensity avoided.\\
Sep~03 & Explainer & Reduced interval reps to cap load spike; HR--pace coupling remained stable in subsequent session.\\
Sep~18 & Motivator & Reframed missed long run; redistributed volume across midweek to sustain adherence.\\
\bottomrule
\end{tabular}
\end{table}

\section{Results}
\label{sec:results}

\subsection{Training Process Trajectories (RQ1)}
Training load followed a gradual build with expected fluctuations due to fatigue and scheduling constraints. Weekly mileage increased from \(\,<10\,\mathrm{km}\) in the initial week to \(>\,35\,\mathrm{km}\) at peak, while session adherence averaged \(70\text{--}80\%\). Intensity distribution was deliberately polarized toward easy running (\(\approx 70\%\) of sessions), with the remainder split across tempo, interval, and long-run sessions in line with the periodized plan.

\begin{figure}[h]
    \centering
    \includegraphics[width=\linewidth]{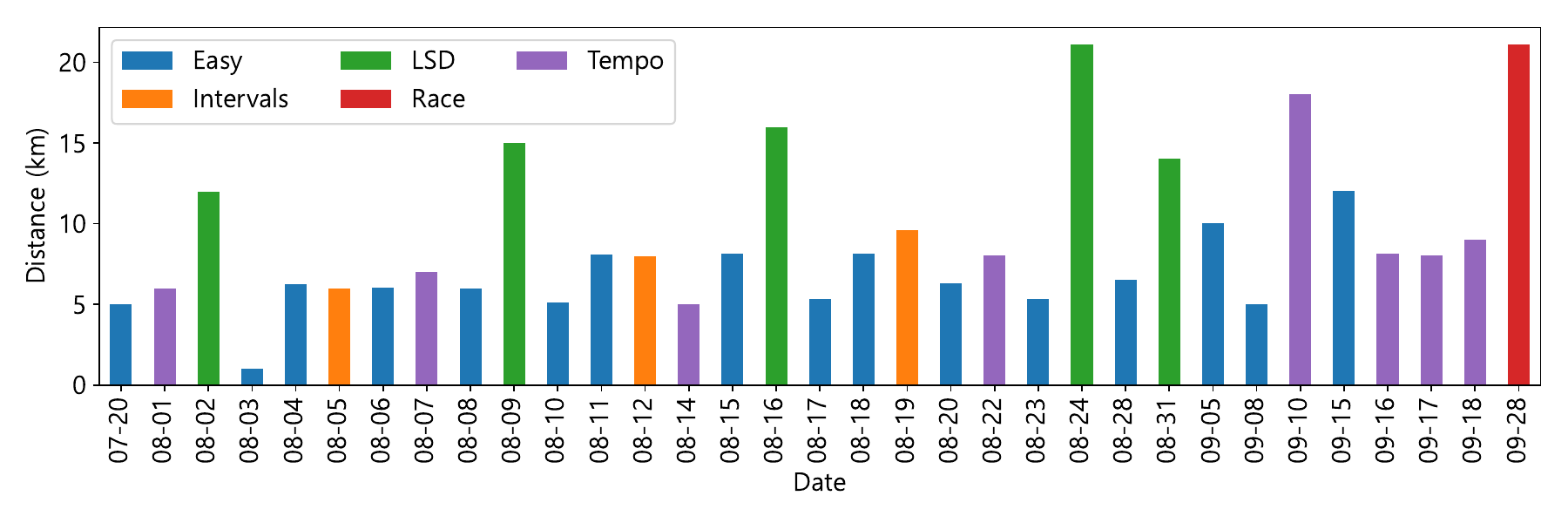}
    \caption{Daily running mileage across the two-month build. Bars represent per-session distance, beginning with the baseline assessment on July 25, 2025.}
    \label{fig:dailymileage}
\end{figure}

\begin{figure}[h]
    \centering
    \includegraphics[width=\linewidth]{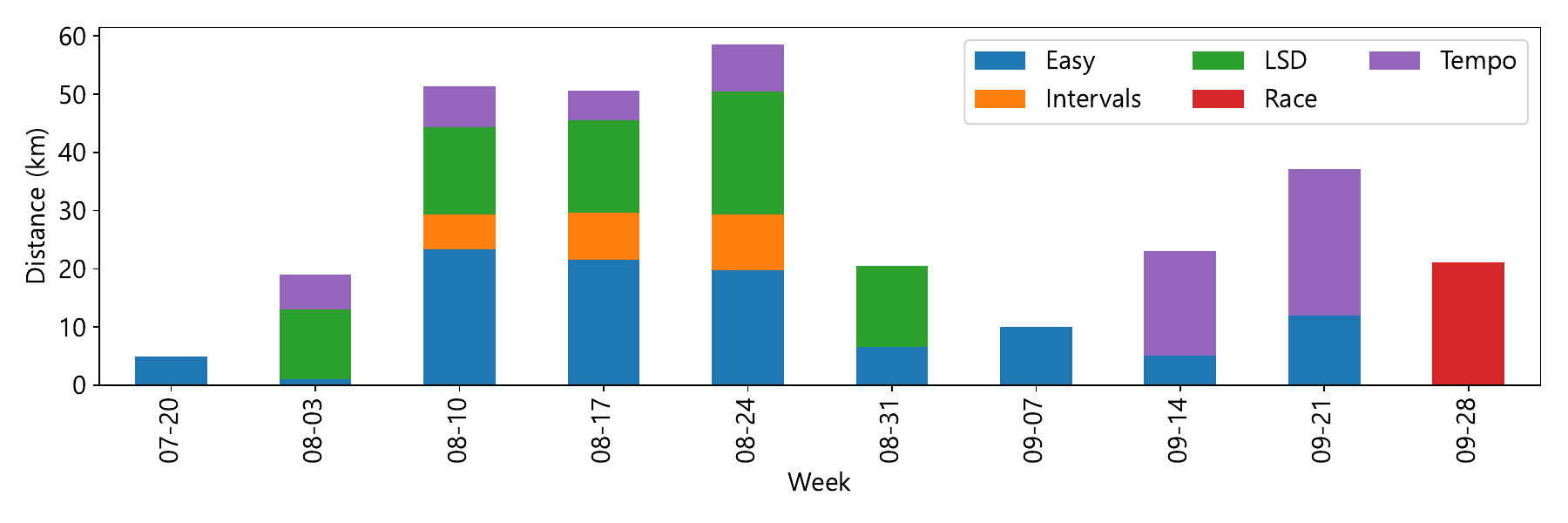}
    \caption{Weekly running mileage. Fluctuations reflect planned load progressions, recovery-driven adjustments, and tapering into the September 28 race.}
    \label{fig:weeklymileage}
\end{figure}

\begin{figure}[h]
    \centering
    \includegraphics[width=\linewidth]{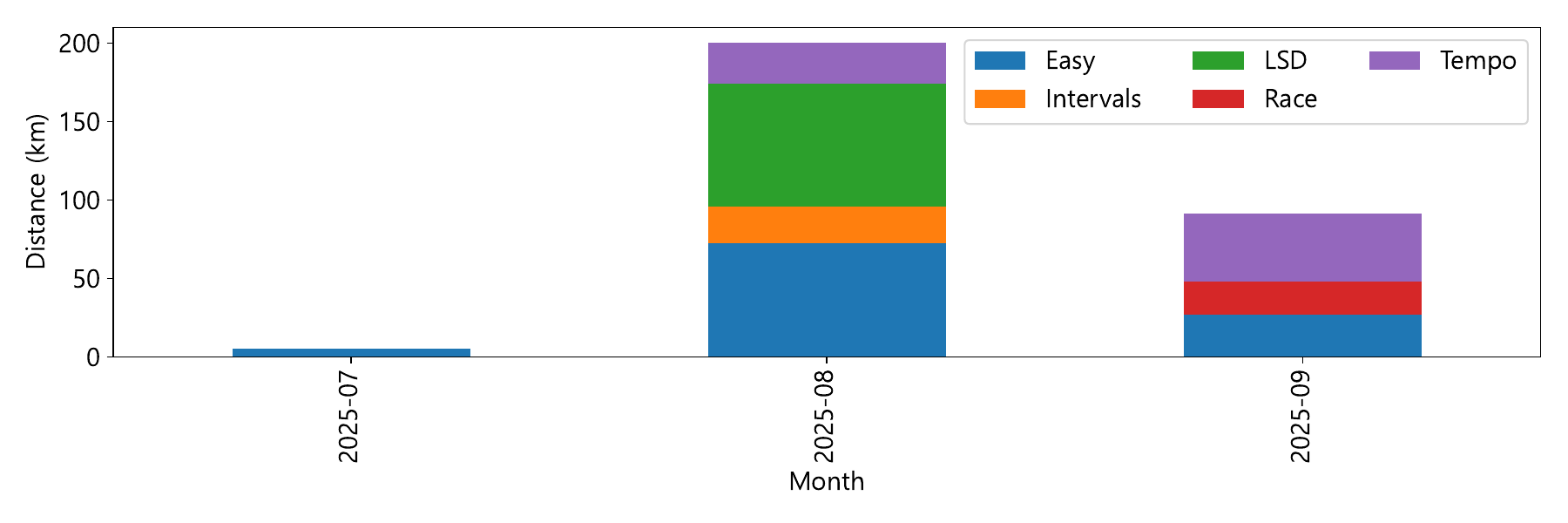}
    \caption{Monthly mileage summary. August accumulated \(\sim 200\,\mathrm{km}\); September totaled \(\sim 100\,\mathrm{km}\) including the half-marathon (21.1\,km).}
    \label{fig:monthlymileage}
\end{figure}

Figures~\ref{fig:dailymileage}--\ref{fig:monthlymileage} visualize daily, weekly, and monthly running mileage, respectively, addressing \textbf{RQ1} at the process level. Training commenced with a baseline run on July 25, 2025 and then adopted a repeating weekly template adapted on demand: \emph{easy/recovery runs} (Mon/Wed/Fri) to consolidate aerobic capacity and promote recovery; \emph{intervals} (Tue) to stimulate maximal oxygen uptake and economy; \emph{tempo runs} (Thu) to target lactate-threshold adaptation; and \emph{long slow distance (LSD)} (Sat) to expand endurance capacity. In practice, this scaffold was flexibly adjusted to accommodate personal schedule, fatigue, and weather.

Accumulated mileage reflected a classic pattern of load build-up followed by taper: approximately \(200\,\mathrm{km}\) were covered in August during base and early progression phases, followed by \(\sim 100\,\mathrm{km}\) in September including race day (21.1\,km). This exposure profile is consistent with periodized endurance paradigms, wherein high proportions of low-intensity running support durable aerobic adaptations, and targeted “point sessions” (intervals, tempo, long runs) provide specific stimuli for VO\(_2\)max, lactate tolerance, and muscular endurance~\cite{Seiler2010,Stoggl2014}.

Prior to this program, the participant—like many recreational runners—assumed that performance hinges primarily on “training hard” (maximizing pace and elevating heart rate). The observed trajectories contradict this misconception: a high share of \emph{easy} sessions enabled sustainable progression and recovery, while strategically placed higher-intensity sessions delivered focused adaptations. The resulting balance between low- and high-intensity work aligns with contemporary recommendations for endurance training distribution~\cite{Seiler2010,Stoggl2014}.


\subsection{Performance Outcomes (RQ1)}
\label{sec:perf}

\begin{figure}[h]
    \centering
    \includegraphics[width=\linewidth]{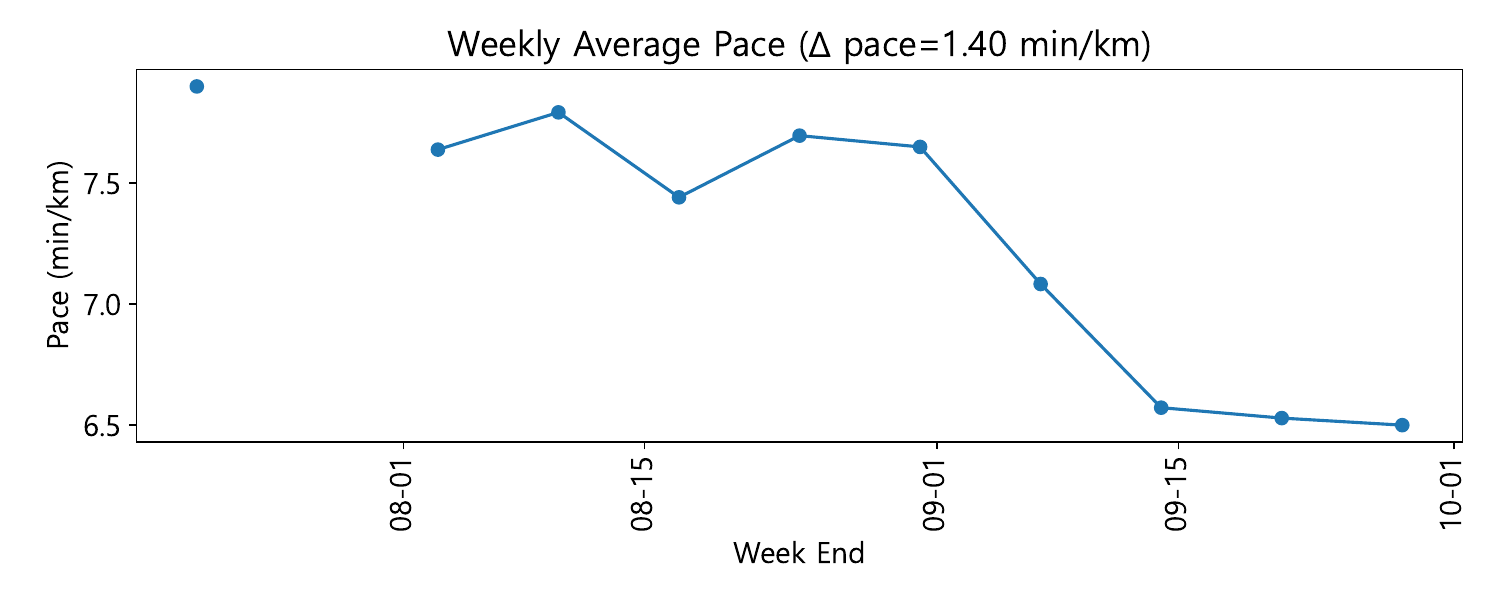}
    \caption{Weekly average pace across the training period (lower is faster). A steady downward trend indicates improved sustainable speed.}
    \label{fig:weeklyavgpace}
\end{figure}

\begin{figure}[h]
    \centering
    \includegraphics[width=\linewidth]{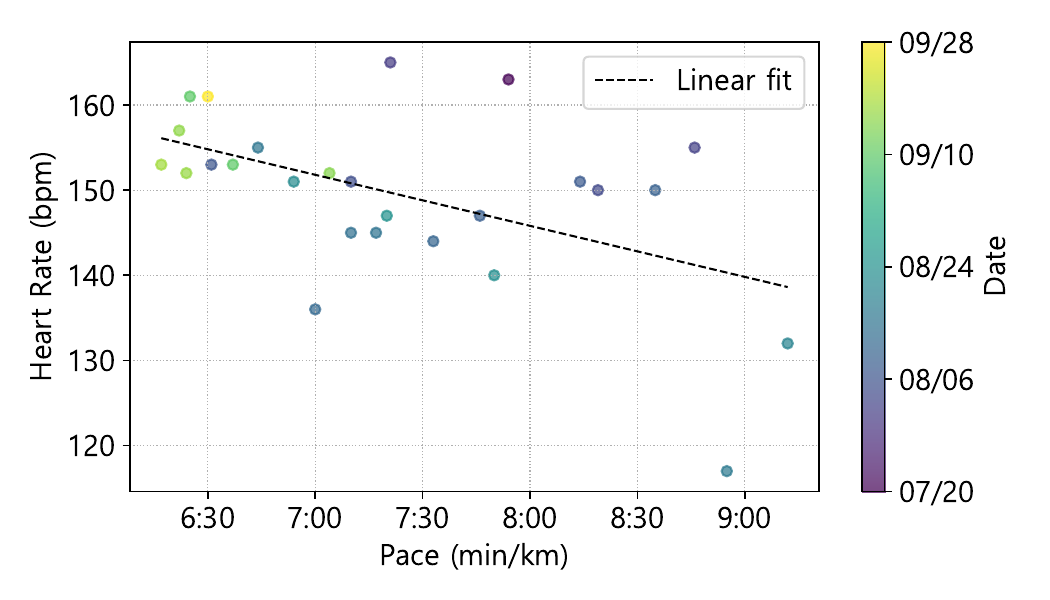}
    \caption{Pace–heart rate coupling across all sessions. Color encodes date (earlier$\to$later), and the dashed line is a within-sample linear fit. Later points shift toward lower HR at a given pace, consistent with improved aerobic economy.}
    \label{fig:pacevsHR}
\end{figure}

\noindent
Pace (min/km) is a primary indicator of running performance. As shown in Figure~\ref{fig:weeklyavgpace}, the weekly average pace improved progressively, reflecting the ability to sustain faster speeds with continued training exposure. Complementarily, Figure~\ref{fig:pacevsHR} shows that sessions later in the block lie below the fitted HR–pace trend observed early on: for the same external load (pace), the internal load (mean HR) declined. This pattern indicates enhanced aerobic efficiency and cardiovascular conditioning. The largest gains occurred in early–mid August, coinciding with the period of highest overall mileage.

It is important to note that these gains occurred while following plans produced in dialogue with the LLM; however, absent a control condition (e.g., independent training without LLM support), we refrain from causal attribution. What the data do show is a clear, time-locked improvement in performance during the two-month program.

\begin{figure}[h]
    \centering
    \includegraphics[width=\linewidth]{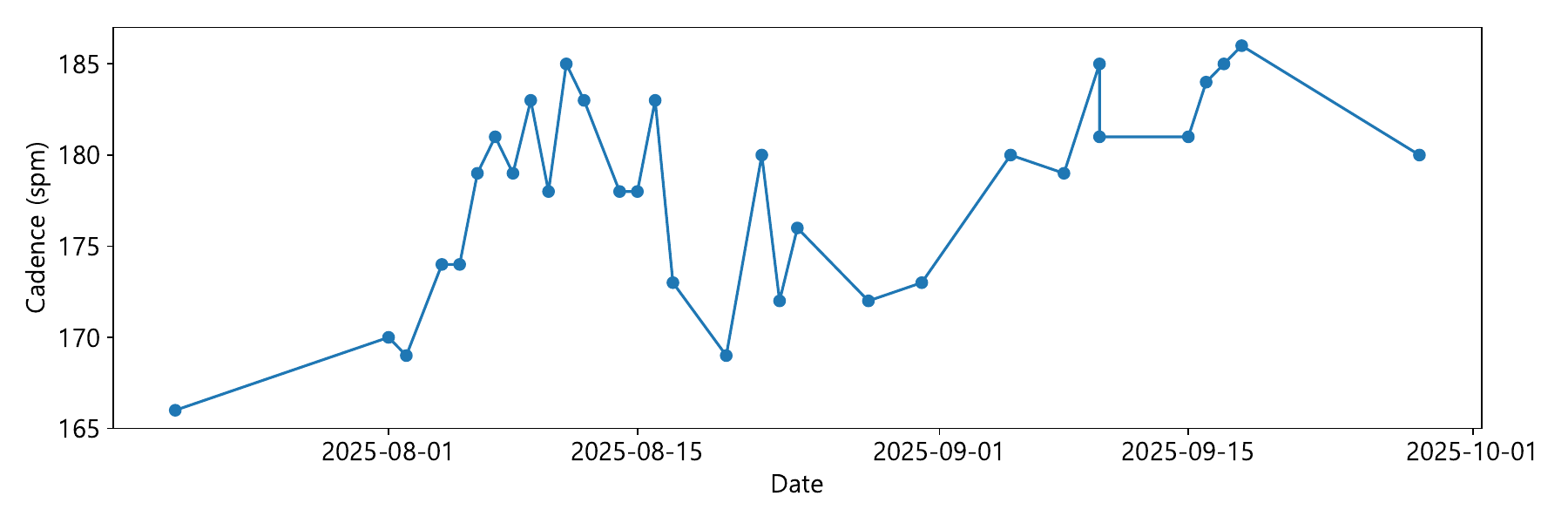}
    \caption{Running cadence (steps per minute, spm) over time. Early sessions clustered near 160\,spm; later sessions show consistent maintenance within the 170–180\,spm target range, aligning with coaching guidance.}
    \label{fig:cadence}
\end{figure}

\noindent
Cadence (spm) is a biomechanical marker of economy. The participant began near 165\,spm, which at a given pace typically implies longer strides, increased braking, and higher joint loading. The LLM recommended progressive cadence elevation toward 170–180\,spm. Figure~\ref{fig:cadence} shows a sustained shift into this range across easy, tempo, interval, and long runs, suggesting not only a numerical increase but also a correction of habitual mechanics toward a more economical, lower-impact stride.

\medskip
\noindent
Beyond external outputs (pace, distance), we quantified cardiovascular efficiency using an \emph{efficiency index (EI)}—the ratio of speed (km/h) to average HR (bpm) for each session:
\[
\mathrm{EI} \;=\; \frac{\mathrm{speed\;(km/h)}}{\mathrm{avg\_HR\;(bpm)}}, 
\qquad 
\mathrm{speed} \;=\; \frac{60}{\mathrm{pace\;(min/km)}}.
\]
Higher EI denotes more speed per unit cardiac cost. We computed daily EI and summarized trends via distance-weighted 7- and 28-day medians.

\begin{figure}[h]
    \centering
    \includegraphics[width=\linewidth]{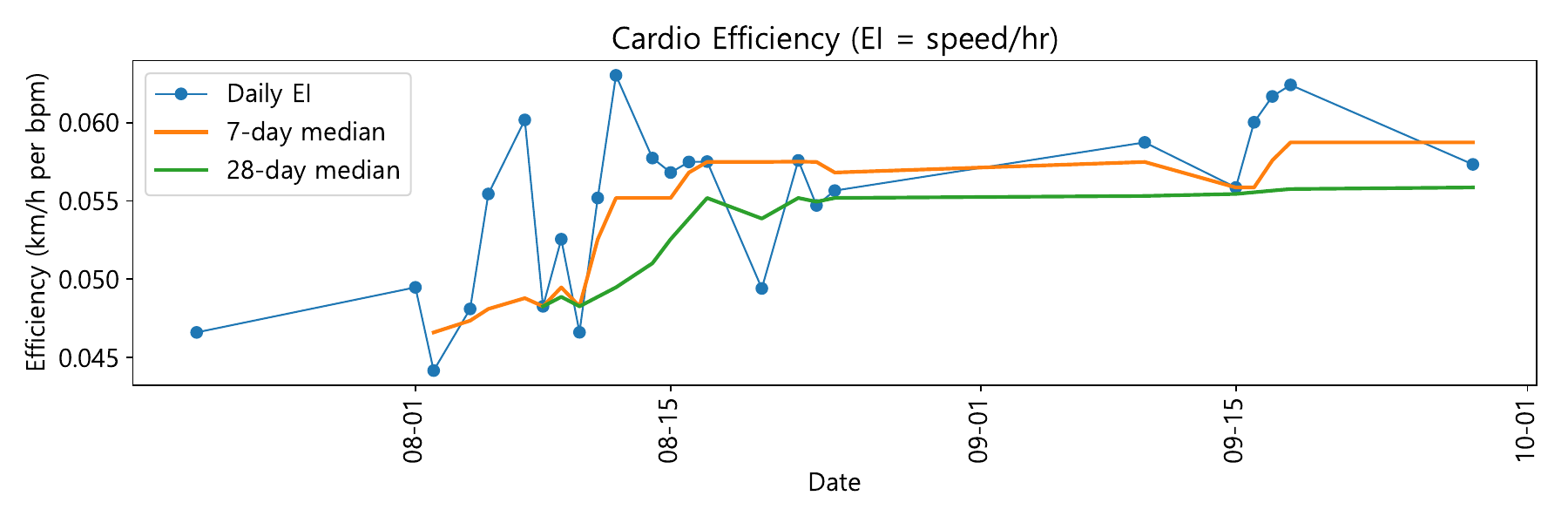}
    \caption{Daily efficiency index (EI) with 7- and 28-day rolling medians. Despite day-to-day variability, both smoothed series trend upward, indicating improved speed per unit cardiac cost.}
    \label{fig:dailyEI}
\end{figure}

\begin{figure}[h]
    \centering
    \includegraphics[width=\linewidth]{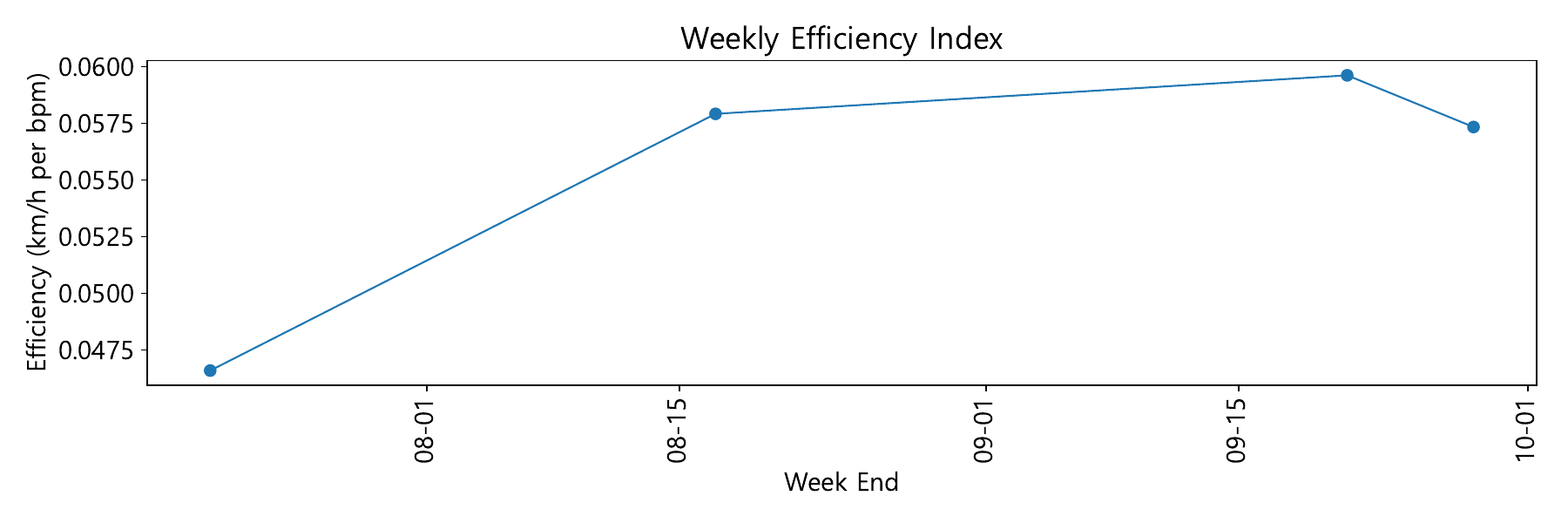}
    \caption{Weekly, distance-weighted EI aggregates. Emphasis on longer sessions shows that efficiency gains extend to sustained training exposure, not only isolated runs.}
    \label{fig:weeklyEI}
\end{figure}

\noindent
Figures~\ref{fig:dailyEI} and~\ref{fig:weeklyEI} corroborate the HR–pace findings: EI increased across the block at both daily and weekly scales, with short-horizon (7-day) and medium-horizon (28-day) summaries moving in concert. Together with Figures~\ref{fig:weeklyavgpace} and~\ref{fig:pacevsHR}, these results indicate concurrent improvements in external performance and internal efficiency during the study period.

\begin{figure}[h]
    \centering
    \includegraphics[width=\linewidth]{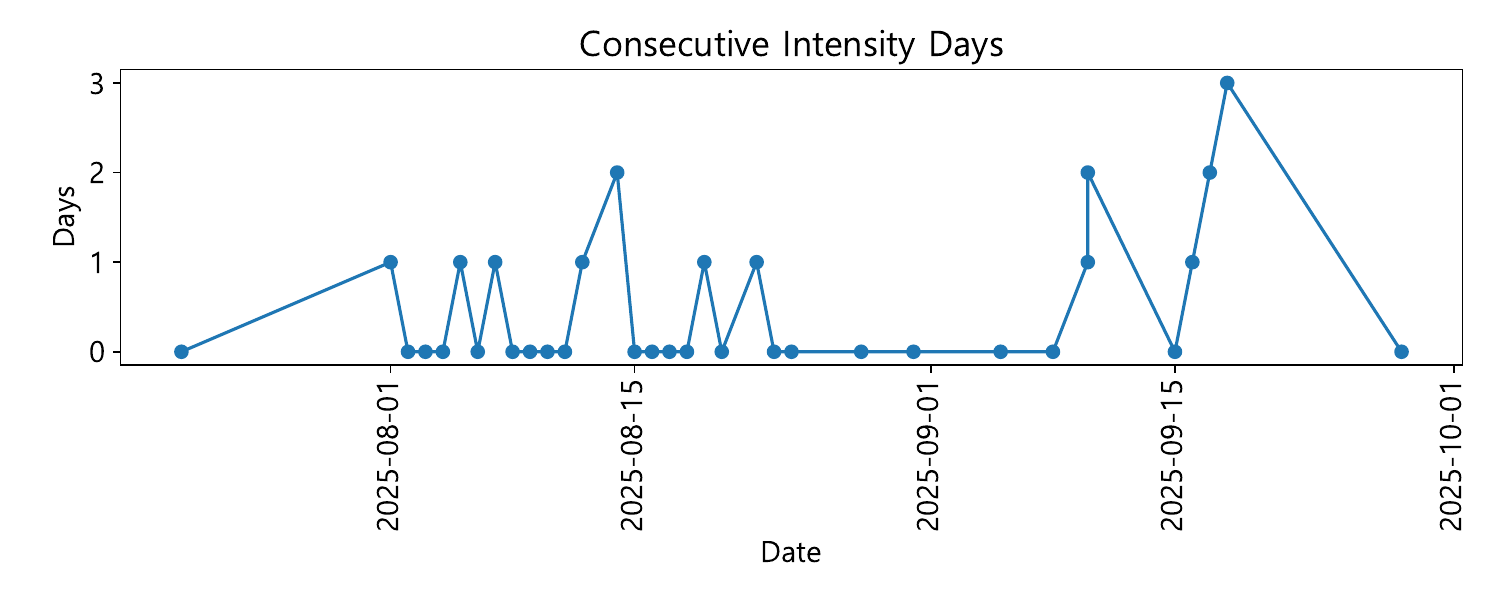}
    \caption{Consecutive intensity days (tempo or interval). Lower values reflect better adherence to load-distribution guidance. Rare spikes correspond to deliberate schedule compressions near pacing-focused microcycles in September.}
    \label{fig:consecutive}
\end{figure}

\noindent
Finally, we examined \emph{consecutive intensity days} as a safety-related process check. The LLM advised separating high-intensity sessions to limit cumulative fatigue. As Figure~\ref{fig:consecutive} shows, back-to-back intensity was uncommon, with rare exceptions when compressing schedules ahead of race preparation. In most weeks, intensity sessions were separated by at least one day, helping maintain balanced load distribution and mitigating overuse risk.






\begin{figure}[h!]
\centering
\begin{mdframed}[backgroundcolor=gray!5, roundcorner=5pt]
\small
\textbf{Q: I feel a mild ``pinching'' discomfort in my right hip during runs. It is not severe, but it is noticeable.}\\[0.25em]
\textbf{A:} A pinching sensation at the front of the hip during running can arise when hip flexors or lateral gluteal stabilizers (e.g., gluteus medius/minimus) underperform, increasing anterior joint stress. Here is a staged response:

\emph{1) Immediate adjustments}
\begin{itemize}\itemsep2pt
  \item \textbf{Form checks:} Reduce overstriding (avoid the foot landing too far ahead of the center of mass); maintain a slight forward lean to unload hip flexors.
  \item \textbf{Warm-up upgrades:} Replace walking lunges with \emph{90/90 hip mobility} (8 reps/side) and targeted hip flexor release (30\,s/side).
  \item \textbf{Cool-down:} Hip-flexor stretch (rear-foot on floor, gentle pelvis drive, 40\,s) and supine figure-four glute stretch (40\,s/side).
\end{itemize}

\emph{2) Short- and mid-term training adjustments}
\begin{center}
\begin{tabular}{p{0.26\linewidth} p{0.30\linewidth} p{0.36\linewidth}}
\toprule
\textbf{Domain} & \textbf{Planned} & \textbf{Adjustment}\\
\midrule
Tempo (8/7) & 7 km @ 7'15'' & Keep 7 km, but if discomfort increases, reduce main tempo to 5 km; add 2 km split into extended WU/CD.\\
Intervals & 400 m $\times$ 5--8 & Halve total reps (to 4) or slow by $\sim$5'/400 m; extend recovery jog from 400 m $\to$ 600 m.\\
Long run (8/9) & 14 km & If symptoms rise, reduce to 12 km and include a 1 km walk intermission.\\
\bottomrule
\end{tabular}
\end{center}

\emph{3) Targeted strength \& mobility}
\begin{itemize}\itemsep2pt
  \item \textbf{Glute med activation:} Side-lying hip abduction 3$\times$15/side.
  \item \textbf{Hip-flexor isometrics:} Split-stance lunge holds 2$\times$30\,s/side.
  \item \textbf{Soft-tissue:} Foam roll quadriceps and iliopsoas $\sim$1\,min each.
\end{itemize}

\textit{Notes:} If discomfort persists or escalates, further reduce intensity/volume and consider clinical assessment. Report post-session status for additional adjustments.
\end{mdframed}
\caption{Example LLM planning under mild right-hip discomfort: near-term form cues, session modifications, and targeted accessory work.}
\label{fig:qa-hip}
\end{figure}

\begin{figure}[h!]
\centering
\begin{mdframed}[backgroundcolor=gray!5, roundcorner=5pt]
\small
\textbf{Q: I fully rested on 8/25 due to being busy and some leg soreness. On 8/26 I completed lower-body + core weight training. Please provide a revised running plan.}\\[0.25em]
\textbf{A:} Given the full rest on 8/25 and the weights session on 8/26, here is an adjusted weekly plan (8/26--9/1) accounting for residual fatigue:

\textbf{8/26 (Tue)}: Weights completed; \emph{no run}. Focus on recovery.\\
\textbf{8/27 (Wed) — Easy Run}: 6--7 km @ 7'30''--8'00''/km, HR 135--145 (Z2).\\
\textbf{8/28 (Thu) — Tempo Run}: WU 1 km $\to$ Main 6 km @ 6'20''--6'30''/km $\to$ CD 1 km; HR 155--165, RPE 6.\\
\textbf{8/29 (Fri) — Recovery jog + Mobility}: 4--5 km @ $\ge$8'30''/km, HR $<$135; mobility + foam roll (knee/hip).\\
\textbf{8/30 (Sat) — Long Run (LSD)}: 16--18 km @ 6'50''--7'10''/km; HR 145--155 (Z2); gels at 8 km and 14 km.\\
\textbf{8/31 (Sun) — Active Recovery}: 4 km jog @ 8'30''/km \emph{or} 40\,min brisk walk; light core + stretching.\\
\textbf{9/1 (Mon) — Rest}: Full rest or very light stretching.

\textit{Weekly total:} $\sim$32--35 km (Wed--Sun), balanced against the weights session to limit joint/hip stress.\\
\textit{Option:} If desired, provide HR-based pacing table for the LSD.
\end{mdframed}
\caption{Example LLM week restructuring after a rest day and a lower-body weights session.}
\label{fig:qa-replan-1}
\end{figure}

\begin{figure}[h!]
\centering
\begin{mdframed}[backgroundcolor=gray!5, roundcorner=5pt]
\small
\textbf{Q: I have a baseball game on Saturday; doing an 18 km LSD that day seems difficult. Can we move it?}\\[0.25em]
\textbf{A:} Yes. We will preserve the macro-structure (intensity distribution) while shifting the LSD earlier:

\textbf{9/1 (Mon) — Recovery / Easy}: 5 km @ 8'30''--9'00''/km, HR $<$135; mobility.\\
\textbf{9/2 (Tue) — Intervals}: WU 2 km; 1000 m $\times$ 6 @ 5'20''--5'30''/km; 400 m jog recovery between reps; CD 1 km; HR 165--175, RPE 7--8.\\
\textbf{9/3 (Wed) — Easy}: 6--7 km @ 7'30'--8'00'/km, HR 135--145.\\
\textbf{9/4 (Thu) — \underline{LSD (moved earlier)}}: 18 km @ 6'50''--7'10''/km, HR 145--155 (Z2); gels at 8 km and 14 km.\\
\textbf{9/5 (Fri) — Recovery + Strength}: 4--5 km recovery jog @ 8'30''/km; lower-body strength (lunges, hip bridge, squats).\\
\textbf{9/6 (Sat) — Baseball day}: Full rest \emph{or} optional 3--4 km very easy jog in the morning.\\
\textbf{9/7 (Sun) — Tempo}: WU 1 km; Main 6 km @ 6'20''--6'30''/km; CD 1 km; HR 155--165, RPE 6.

\textit{Weekly total:} $\sim$45 km. Intensity order (Tue: intervals, Thu: LSD, Sun: tempo) preserves stimulus separation while avoiding the Saturday conflict.
\end{mdframed}
\caption{Example LLM calendar-aware rescheduling: moving the LSD to accommodate a Saturday baseball game while preserving intensity separation.}
\label{fig:qa-replan-2}
\end{figure}

\subsection{LLM as Planner: Constraint-Aware Micro/Meso Structuring (RQ2)}
\label{sec:planner}

Across routine and exceptional circumstances, the LLM consistently operated as a \emph{planner}, integrating constraints and restructuring sessions to preserve load distribution, safety, and goal alignment. The following instances illustrate how micro-level edits respected mesocycle intent.

\paragraph{(1) Symptom-aware adjustments.}
When mild right-hip discomfort emerged (Figure~\ref{fig:qa-hip}), the model linked symptoms to actionable modifications: immediate form cues (reduce overstriding; slight forward lean), warm-up/cool-down augmentation, short-term reductions in tempo/interval stress, and targeted accessory work (gluteus medius activation; hip-flexor isometrics). This \emph{symptom-to-structure} mapping modulated the microcycle without derailing the mesocycle’s specificity.

\paragraph{(2) Recovery-informed weekly restructuring.}
Following an unplanned rest day and a lower-body weights session (Figure~\ref{fig:qa-replan-1}), the planner redistributed intensity across the week (easy $\rightarrow$ tempo $\rightarrow$ recovery $\rightarrow$ LSD $\rightarrow$ active recovery), increasing total mileage only within conservative bounds. The sequence preserved intended stimuli (threshold and long endurance) while spacing stressors to mitigate cumulative fatigue.

\paragraph{(3) Calendar-aware rescheduling with intensity separation.}
Given a Saturday conflict (baseball game), the model advanced the LSD to Thursday and shifted tempo to Sunday (Figure~\ref{fig:qa-replan-2}), maintaining separation among interval, long, and tempo sessions. This exemplifies pragmatic \emph{plan--calendar co-optimization} that protects polarization principles while honoring external commitments.

\paragraph{(4) Guardrails and progressive overload.}
Across scenarios, the planner role applied implicit guardrails—e.g., halving interval volume when symptomatic, extending recoveries, constraining LSD to Zone~2—and upheld progressive overload at the mesocycle level (weekly totals $\sim$32--45\,km with long-run and tempo anchors). These controls lowered injury risk while maintaining directional progression.

\medskip
\noindent\textbf{Implications.}
The planner translated high-level goals into feasible schedules under evolving constraints (symptoms, fatigue, calendar), preserving stimulus spacing, gradual progression, and specificity. Where disruptions occurred, it proposed concrete, safety-aware alternatives rather than cancelations. As discussed in Section~\ref{sec:limits}, residual limitations (e.g., missing in-run HR/cadence/posture signals; reliance on user-reported context) motivate tighter multimodal integration and explicit safety logic (caps on weekly load changes; automated flags for back-to-back intensity) in future systems.

\begin{figure}[h!]
\centering
\begin{mdframed}[backgroundcolor=gray!5, roundcorner=5pt]
\small
\textbf{Q: What static stretches should I do for 5 minutes after running?}\\[0.25em]
\textbf{A:} Here are six static stretches you can complete in about five minutes. Hold each for 40--45 seconds with 5--10 seconds between movements, maintaining relaxed breathing:
\begin{enumerate}\itemsep2pt
  \item \textbf{Hamstring stretch (seated forward fold):} Sit with legs extended, hinge at the hips with a neutral spine, and reach toward your feet to feel tension along the posterior thigh.
  \item \textbf{Quadriceps stretch (standing heel-to-glute):} Stand and pull one ankle toward the glute. Keep knees together and feel tension across the anterior thigh.
  \item \textbf{Calf stretch (wall-assisted):} With hands on a wall, step one leg back and drive the heel into the floor while gently bending the front knee.
  \item \textbf{Hip flexor stretch (static lunge):} Step forward into a deep lunge and gently push the pelvis forward to stretch the hip flexors of the rear leg.
  \item \textbf{Glute stretch (supine figure-four):} Lying on your back, cross one ankle over the opposite knee and pull the thigh toward the chest to target the glutes.
  \item \textbf{IT band/lateral hip stretch (standing cross-over):} Cross one leg over the other and lean toward the back leg, maintaining balance with a wall or chair if needed.
\end{enumerate}
\textit{Tips:} Keep breathing natural, move only to a comfortable stretch (not pain), and adjust depth as needed.
\end{mdframed}
\caption{Example LLM response: five-minute post-run static stretching routine.}
\label{fig:qa-stretching}
\end{figure}

\begin{figure}[h!]
\centering
\begin{mdframed}[backgroundcolor=gray!5, roundcorner=5pt]
\small
\textbf{Q: Why is heart-rate (HR) control prioritized? Wouldn't training near my limit push my performance faster?}\\[0.25em]
\textbf{A:} Early in base building (first 4--6 weeks), \textit{Zone~2} (approximately 65--75\% of maximal HR) is prioritized for three reasons:
\begin{enumerate}\itemsep2pt
  \item \textbf{Efficient energy-system development:} Z2 improves fat oxidation and mitochondrial density; jumping straight to high intensity biases anaerobic pathways and accumulates fatigue quickly.
  \item \textbf{Recovery and injury prevention:} Frequent Z4--Z5 work elevates neuromuscular and joint stress, lengthens recovery, and increases overuse risk; Z2 volume permits consistency with manageable recovery.
  \item \textbf{Progressive overload:} A robust aerobic base enables later gains when threshold/VO$_2$max sessions are introduced.
\end{enumerate}
\textbf{When to add intensity?} From week~2 onward, introduce \textit{one} weekly threshold (Z3) or VO$_2$max (Z4) session and periodize monthly (e.g., 3 weeks up, 1 week down). A pragmatic distribution is: weeks 1--4 emphasize Z2 volume, weeks 5--7 expand Z3/Z4, and final 2 weeks taper while preserving some intensity.
\end{mdframed}
\caption{Example LLM explanation: rationale for prioritizing HR-controlled base training before higher-intensity work.}
\label{fig:qa-hr}
\end{figure}

\begin{figure}[h!]
\centering
\begin{mdframed}[backgroundcolor=gray!5, roundcorner=5pt]
\small
\textbf{Q: Isn't my weekly mileage too low?}\\[0.25em]
\textbf{A:} Early mileage may feel modest, but base building prioritizes (i) injury prevention and reliable recovery, (ii) gradual progression (5--10\% weekly increases), and (iii) balanced intensity distribution (interval/tempo/long/easy). At this stage, 20--25\,km/week is an appropriate range for safe adaptation.\\[0.35em]
\textbf{How to scale next week if recovery is good:}
\begin{center}
\begin{tabular}{p{0.38\linewidth} p{0.50\linewidth}}
\toprule
\textbf{Before} & \textbf{After (suggestion)}\\
\midrule
Mon: 6\,km easy & Mon: 6\,km $+$ 3\,km easy (total 9\,km)\\
Wed: Intervals & Wed: Intervals (unchanged)\\
Fri: 6\,km tempo & Fri: 6\,km tempo (unchanged)\\
Sat: Strength $+$ 5\,km easy & Sat: Strength $+$ 7\,km easy\\
Sun: 12\,km long run & Sun: 14\,km long run\\
\bottomrule
\end{tabular}
\end{center}
\textit{Notes:} Additional easy run $\sim$9$'$30$''$/km at 65--70\% MHR; increase long run by 1--2\,km while monitoring HR and RPE. This raises weekly volume toward 25--30\,km with low risk.
\end{mdframed}
\caption{Example LLM guidance: justification for modest base mileage and a safe progression template.}
\label{fig:qa-mileage}
\end{figure}

\begin{figure}[h!]
\centering
\begin{mdframed}[backgroundcolor=gray!5, roundcorner=5pt]
\small
\textbf{Q: What is the ``toktok drill''?}\\[0.25em]
\textbf{A:} A cadence-focused drill emphasizing short, quick steps and minimal ground-contact time to raise leg turnover toward 170--180\,spm.\\[0.35em]
\textbf{How to perform:}
\begin{itemize}\itemsep2pt
  \item \textbf{Setup:} Track, flat path, or treadmill.
  \item \textbf{Posture:} Upright torso, relaxed shoulders, slight knee flexion, elastic ankles; arm swing $\sim$90$^\circ$.
  \item \textbf{Action:} In place or moving lightly, ``tap'' the ground quickly with midfoot, keeping stride very short; 30\,s on, 30\,s easy jog or walk, repeat 4--6 sets.
\end{itemize}
\textbf{Benefits:} Higher cadence, better elastic recoil, reduced overstriding and impact loading, and more stable pacing at target cadence.
\end{mdframed}
\caption{Example LLM instruction: cadence-oriented ``toktok'' drill for economical stride mechanics.}
\label{fig:qa-toktok}
\end{figure}

\subsection{LLM as Explainer: From Prescriptions to Didactic Rationales (RQ2)}
\label{sec:explainer}

Beyond supplying session prescriptions, the LLM recurrently acted as an \emph{explainer}, attaching concise rationales and step-by-step instructions that supported understanding, adherence, and self-efficacy. We observed three complementary didactic functions.

\paragraph{(1) Conceptual grounding of training decisions.}
As illustrated in Figures~\ref{fig:qa-hr} and~\ref{fig:qa-mileage}, the model contextualized base-building priorities (e.g., a Zone~2 emphasis) in terms of energy-system development, recovery management, and progressive overload. This transformed otherwise opaque directives (e.g., “run slower, run more”) into transparent trade-offs (mitochondrial adaptation, injury risk, taper dynamics), enabling acceptance of conservative prescriptions without the perception of undertraining.

\paragraph{(2) Procedural instruction for skill components.}
When queried about post-run routines or cadence drills (Figures~\ref{fig:qa-stretching} and~\ref{fig:qa-toktok}), the LLM provided operational details—set/hold times, body cues, and common faults to avoid—that could be executed immediately. These micro-protocols (e.g., 40–45\,s static holds; 30\,s on/30\,s off cadence bouts for 4–6 sets) reduced friction between intention and action, particularly for a novice unfamiliar with ancillary work.

\paragraph{(3) Explanations that scaffold adaptive planning.}
In response to concerns about “low” mileage (Figure~\ref{fig:qa-mileage}), the LLM coupled justification (injury risk, recovery constraints) with concrete progression steps (adding an easy run; extending the long run by 1–2\,km). Pairing the \emph{why} with the \emph{how} facilitated safe load increases while preserving the participant’s sense of agency.

\medskip
\noindent
\textbf{Implications.}
The explainer role made the logic of training explicit at two time scales: (i) \emph{within-week}—how to execute a stretch, interval set, or drill—and (ii) \emph{across weeks}—why base-first periodization and gradual mileage increases sustain long-term gains. In combination with planner and motivator roles, these didactic interactions helped bridge the gap between generic guidance and individualized, actionable coaching. At the same time, explanatory adequacy remained bounded by available data (e.g., absent HR or cadence during a session), suggesting that tighter multimodal integration (wearable HR/cadence, sleep summaries) would further refine and personalize future explanations (cf.\ Section~\ref{sec:limits}).

\subsection{LLM as Motivator: Limits of Passive, User-Initiated Support (RQ2)}
\label{sec:motivator}

While the agent provided encouraging language when prompted, its motivational impact was fundamentally bounded by a \emph{passive, user-initiated} interaction model.

\paragraph{(1) Opaque motivational state without explicit disclosure.}
The agent could not infer waning motivation unless the participant stated it explicitly. In practice, periods of low engagement (busy schedules, fatigue, dampened affect) were invisible because no behavioral or physiological signals were streamed automatically. As a result, the system lacked a basis for timely prompts (e.g., reframing a missed session, suggesting a lighter substitute) precisely when motivation was at risk.

\paragraph{(2) Asymmetric contact: help least available when most needed.}
During high-motivation phases, the participant executed plans diligently and rarely requested motivational content; the LLM’s role collapsed to confirmation and minor edits. Conversely, during low-motivation weeks the participant often \emph{did not initiate} a chat at all, precluding any intervention. This asymmetry highlights a structural limitation of pull-based designs: when support is most needed, the channel is least likely to be opened.

\paragraph{(3) Ad hoc encouragement vs.\ proactive, just-in-time support.}
Because encouragement was typically triggered by explicit requests, support appeared \emph{ad hoc} rather than continuous or context-aware. The participant perceived a gap relative to human coaching, where coaches proactively check in after missed sessions, detect early signs of dropout, and reframe short-term goals to preserve adherence.

\medskip
\noindent
\textbf{Implications.}
Effective digital motivation likely requires a shift from passive, user-pull interaction to a \emph{proactive, sensing-enabled} model. Two design directions emerge:
(i) \emph{Lightweight check-in scaffolds} (scheduled micro-prompts; push notifications after missed/truncated sessions; brief weekly reflection forms) to elicit motivational state even when the user would not open the chat; and
(ii) \emph{Multimodal state estimation} (calendar load, sleep summaries, missed-run patterns, HR variability, step counts) that triggers \emph{just-in-time} interventions (short replacement workouts, reframing messages, commitment reminders) without requiring explicit requests.
Integrating such mechanisms—with opt-in consent and privacy controls—appears necessary to close the motivational gap observed here and to sustain adherence in real-world training.

\section{Limitations of the Current Approach and Design Directions for Next-Generation Systems}
\label{sec:limits}

This case study shows that an LLM—engaged primarily through text and retrospective logs—can support planning, explanation, and some adherence scaffolding during a two-month half-marathon build. At the same time, observed benefits were bounded by structural limitations in sensing, feedback channels, motivation, and model specialization. We distill these constraints and outline design directions for systems aspiring to function as safe, effective, and personalized virtual coaches in real-world training.

\subsection{Insufficient Modalities and Lack of Real-Time Context (Human $\to$ LLM)}
\label{subsec:modalities-realtime}

A core constraint was the \emph{post hoc}, text-first nature of most interactions. Although daily logs (date, distance, pace, occasionally HR/cadence) enabled high-level plan adaptation, they were insufficient for \emph{in-session} decision-making. During workouts, the coach had no direct visibility into form (pose, overstriding, asymmetries), breathing patterns, transient HR drift, heat stress, or terrain-induced load. As a result, the system could not (i) detect technique faults as they emerged, (ii) issue timely safety prompts (e.g., back off when HR spikes under heat), or (iii) exploit opportunities for micro-adjustments (e.g., shorten recoveries when HR stabilizes quickly).

\textbf{Design implication.}
Future systems should ingest \emph{multimodal, low-latency} signals: on-body sensors (PPG/ECG HR, IMU for cadence/impact, footpods for ground-contact time), phone/glasses cameras for \emph{on-device} pose estimation, and environmental telemetry (temperature, humidity, grade). A lightweight \emph{streaming fusion} layer can map these to interpretable states (e.g., ``probable overstriding,'' ``rising aerobic decoupling,'' ``heat-adjusted HR target''). To close the loop \emph{within} a run, coaches must operate under tight compute/power budgets—arguing for \emph{edge inference} with intermittent cloud sync. Architecturally: (i) \emph{event-driven} pipelines that trigger micro-interventions when thresholds are crossed (e.g., HR–pace decoupling $>$5\%); (ii) \emph{graceful degradation} when sensors are missing; and (iii) explicit \emph{uncertainty handling} so the agent can act (low risk) or query (high ambiguity).

\subsection{Richer, Embodied Feedback Channels (LLM $\to$ Human)}
\label{subsec:feedback-outbound}

Instructions were delivered almost exclusively as text, leaving a gap between \emph{verbal prescriptions} and \emph{embodied execution}. For motor skills—cadence changes, foot strike, hip control—language-only guidance imposes cognitive load and risks misinterpretation; post-run explanations cannot remediate within-run faults that accumulate fatigue and load.

\textbf{Design implication.}
Coaching outputs should be \emph{multimodal and situated}: short \emph{audio} cues (metronome/beat-synced prompts) for cadence; \emph{haptic} cues on wearables for pace/zone drift; and \emph{visual} overlays (phone or glasses) for pose targets with traffic-light semantics (``stable knee track,'' ``reduce overstride''). To avoid overload, feedback must be \emph{bandwidth-aware}: prioritize one cue at a time, escalate as risks grow, and throttle prompts during high-RPE segments. Between sessions, brief video exemplars or stick-figure replays (from IMU/pose data) can support actionable reviews. Outbound messages should include \emph{rationale snippets} and \emph{expected sensations} (``hips should feel engaged; breathing steady'') to help users map text to bodily cues.

\subsection{Motivation as a First-Class Design Target}
\label{subsec:motivation}

A recurring limitation was dependence on \emph{user-initiated} contact for motivational support. When intrinsically motivated, the participant executed plans with minimal coaching; during busy or low-affect periods, the participant often did not open the chat at all—precluding timely reframing or adherence rescue. Thus, the current pull-based interaction fails precisely when motivational interventions are most needed.

\textbf{Design implication.}
Next-generation coaches should (i) \emph{proactively} sense motivational risk and (ii) deliver \emph{just-in-time} interventions. Signals can include missed/shortened runs, reduced step counts, poor sleep, rising HR at fixed paces, or calendar load spikes. With explicit consent, \emph{check-in scaffolds} (micro-surveys, weekly reflection prompts) and \emph{push notifications} after detected lapses can elicit state even when the user would not initiate chat. Behavioral design elements—\emph{gamification} (streaks with compassionate resets), \emph{social accountability} (opt-in micro-groups), and \emph{commitment devices} (pre-committed time slots, soft pledges)—should be available as optional layers. Motivational prompts must remain \emph{non-coercive}, with user controls for frequency, modality, and quiet hours.

\subsection{Security, Personalization, and Model Specialization}
\label{subsec:security-model}

We intentionally used a general-purpose LLM without domain-specific training or privileged data flows. This minimized setup but limited personalization, robustness, and transparency. A generic model cannot, by default, encode athlete-specific constraints (injury history, HR zones, gait idiosyncrasies) as a persistent, safety-aware state machine.

\textbf{Design implication.}
We envision a \emph{coach stack} with:
(1) a \emph{persistent athlete model} (goals, zones, injury flags, historical load) backed by explicit memory/schema;
(2) \emph{safety logic} and guardrails (caps on weekly load increases; flags for back-to-back intensity; heat/humidity adjustments) that the LLM must respect;
(3) \emph{privacy-preserving learning}: \emph{federated learning} to adapt shared components without centralizing raw sensor data; \emph{on-device adapters/fine-tuning} for personal style and constraints; and \emph{test-time adaptation} to recalibrate HR–pace relations under acute changes (heat, altitude, illness);
(4) \emph{auditable reasoning} via structured rationales (citing guardrail rules or load metrics), supporting user trust and clinician review where relevant.
Security controls should prioritize local retention by default, differential privacy for aggregated updates, and user-facing dashboards for consent, retention windows, and data export.

\paragraph{From prototype to platform.}
Operationally, these directions suggest a layered architecture:
\emph{Sensing layer} (wearables, phone, camera) $\to$
\emph{Edge inference} (pose, HRV, decoupling) $\to$
\emph{State \& guardrail layer} (athlete profile, rules, uncertainty) $\to$
\emph{Dialogue/feedback layer} (LLM with audio/haptic/visual outputs).
This separation enables incremental deployment (e.g., start with HR+pace; later add pose) and clearer safety testing.

\medskip
\noindent
\textbf{Key Takeaways.}
\begin{itemize}\itemsep4pt
  \item \textbf{Multimodal, real-time sensing.} Replace text-only, post hoc logs with low-latency signals (HR/HRV, IMU, footpods, on-device pose, environment). Use \emph{edge inference} and event-driven triggers (e.g., HR–pace decoupling $>$5\%) to enable within-run micro-adjustments under power/compute constraints.
  \item \textbf{Embodied, bandwidth-aware feedback.} Go beyond text by combining short \emph{audio} (cadence cues), \emph{haptics} (zone/pace drift), and \emph{visual} overlays (traffic-light pose feedback). Throttle prompts during high-RPE segments, and pair guidance with brief rationale snippets and expected sensations.
  \item \textbf{Proactive motivation and JIT interventions.} Move from passive, user-initiated chats to sensing-enabled check-ins: detect risk via missed runs, sleep/HRV trends, step-count dips, and calendar spikes; trigger opt-in push prompts, micro-surveys, and short replacement workouts to prevent dropout.
  \item \textbf{Secure personalization with explicit guardrails.} Maintain a persistent athlete model (goals, zones, injuries) and safety logic (weekly load caps, flags for back-to-back intensity, heat/humidity adjustments). Adapt privately via federated learning and on-device adapters; support \emph{auditable} rationales for trust and review.
\end{itemize}

\medskip
\noindent
\textbf{Linking Results to Research Questions.}
The findings map directly onto the research questions in Section~\ref{sec:design}.
For \textbf{RQ1} (training outcomes under LLM-assisted planning), analyses of pace trajectories, mileage growth, cadence improvements, and efficiency indices (Figures~\ref{fig:weeklymileage}--\ref{fig:weeklyEI}) show measurable aerobic and biomechanical adaptation over two months.

For \textbf{RQ2} (decision-level support), vignettes and role analyses demonstrate effective \emph{planner} and \emph{explainer} functions, while the \emph{motivator} role was limited by passive initiation.

For \textbf{RQ3} (functional gaps), the observed limitations—absence of real-time sensor integration, lack of multimodal feedback channels, insufficient motivational detection, and minimal safety-aware guardrails—crystallize requirements for next-generation systems. Together, these results motivate a roadmap toward context-aware, personalized, and sustainable AI coaching in endurance training.

\section{Threats to Validity}
\label{sec:threats}
Threats include the single-participant design, self-report biases, external confounders (weather, course), and measurement error from consumer-grade devices. We mitigate these by releasing anonymized logs, pre-defining metrics, and clearly separating pre- and post-event comparisons.
\section{Conclusion}
\label{sec:conclusion}
This study investigated what a general-purpose LLM can—and cannot—do as a running coach over a bounded, event-focused window. Using ordinary consumer tools and predominantly text-first interaction, the agent supported planning (micro/mesocycles and schedule repair), explanation (energy systems, taper rationales, drill instructions), and selective motivation (reframing and encouragement upon request). The participant progressed from $\sim$2\,km at $7'54''/\mathrm{km}$ to completing a half-marathon at $\sim$${6'30''/\mathrm{km}}$, accompanied by improvements in cadence (from $\sim$160\,spm toward 170–180\,spm), favorable shifts in the pace–HR relationship, and a steadily rising efficiency index. Although the single-subject design and absence of a control arm preclude causal claims, the combination of process traces and quantitative markers demonstrates that LLM-mediated planning and reflection can align with safe, measurable progress over weeks, not merely days.

Equally important are the boundaries we observed. First, \textbf{modality and real-time gaps} limited in-run safety and technique coaching: without pose, cadence fidelity, or heat-adjusted HR context, guidance arrived primarily before or after sessions. Second, \textbf{outbound feedback channels} were mismatched to embodied skill learning: text alone struggled to convey motor cues as effectively as audio, haptics, or visualizations. Third, \textbf{motivation remained user-pull}: when engagement waned, the chat channel often remained unopened, blunting adherence rescue. Fourth, \textbf{model specialization and guardrails} were implicit rather than explicit: persistence of athlete state, caps on weekly load changes, and calendar-aware risk flags were emulated conversationally rather than enforced systematically.

From these findings, we derive a \emph{design program} for next-generation coaching systems: (i) a persistent, longitudinal athlete model (goals, HR zones, injury flags, historical load) coupled to explicit, auditable safety logic (weekly load caps, back-to-back intensity checks, heat/humidity adjustments); (ii) multimodal, \emph{on-device} sensing and inference (HR/HRV, IMU, optional pose) that enable \emph{near–real-time} cues under tight power/latency budgets; (iii) bandwidth-aware outbound feedback (audio metronomes, haptics for zone drift, lightweight visual overlays) with concise rationales that connect sensations to prescriptions; (iv) proactive, privacy-preserving motivation scaffolds (opt-in check-ins, streaks with compassionate resets, small-group accountability) triggered by consented signals (missed runs, sleep, calendar load); and (v) privacy-advancing personalization via federated learning, on-device adapters, and test-time adaptation to maintain performance without centralizing raw data.

In sum, this work contributes (1) a richly annotated account of LLM-assisted, event-driven endurance preparation, (2) an analysis framework that maps \emph{planner/explainer/motivator} roles to concrete training decisions and outcomes, and (3) a set of actionable requirements that translate observational gaps into system features. We view these as pragmatic groundwork for controlled, multi-participant studies with stronger baselines, objective field tests, and explicit safety evaluations—steps needed to move LLM coaching from promising prototypes to trustworthy, everyday training companions.

\begin{figure}[h]
    \centering
    \includegraphics[width=.7\linewidth, angle=180]{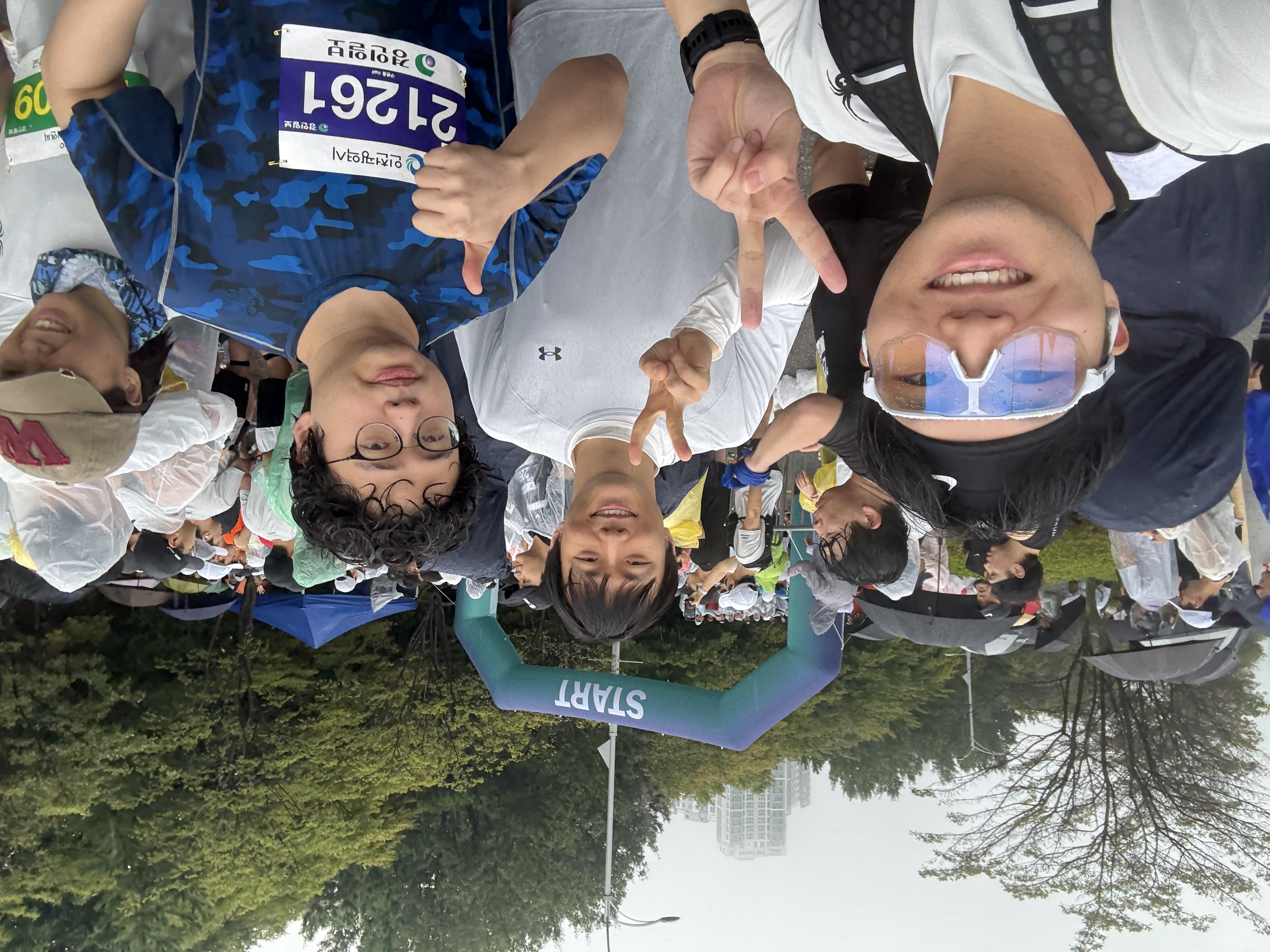}
    \caption{Picture of Kichang Lee (left), YoonTae Kim (center), and Dr.\ Bonhong Koo (right) at the Incheon Songdo International Marathon 2025.}
    \label{fig:pic}
\end{figure}
\begin{acks}
Running offered me—and, I hope, many fellow graduate students navigating physical and mental strain—a concrete way to rebuild health in both body and mind. I wish that more people may discover, through running, new forms of joy and well-being. I am especially grateful to my parents for enabling my return to an active life despite a long orthopedic history, including five surgeries on my right knee with residual issues involving the tibia, ACL, PCL, and meniscus. Their support made it possible to create the conditions and regain the capacity to train again. I also thank my training partners at Yonsei University, Sungmin Lee and YoonTae Kim, and my senior colleague Dr.\ Bonhong Koo, for preparing together, sharing the miles, and running the race with me.
\end{acks}


\bibliographystyle{ACM-Reference-Format}
\bibliography{reference}

\clearpage

\end{document}